\documentstyle[11pt,aaspp4,flushrt,tighten,psfig]{article}  % prep 1 colonna
\begin{document}

\title{Detection of the Entropy of the Intergalactic Medium: Accretion
Shocks in Clusters, Adiabatic Cores in Groups}

\author{Paolo Tozzi\altaffilmark{1}} 
\affil{Department of Physics and Astronomy, The Johns
Hopkins University, Baltimore, MD 21218}  
\author{Caleb Scharf} 
\affil{Space Telescope Science Institute, 3700 San Martin Drive, 
Baltimore, MD 21210}
\and
\author{Colin Norman\altaffilmark{2}} 
\affil{Department of Physics and Astronomy, The Johns
Hopkins University, Baltimore, MD 21218.}  

\altaffiltext{1}{Osservatorio Astronomico di Trieste, via Tiepolo 11, I-34131
Trieste, Italy}
\altaffiltext{2}{Space Telescope Science Institute, 3700 San Martin Drive, 
Baltimore, MD 21210}

\begin{abstract}

The thermodynamics of the diffuse, X-ray emitting gas in clusters of
galaxies is linked to the entropy level of the intra cluster
medium. In particular, models that successfully reproduce the
properties of local X-ray clusters and groups require the presence of
a minimum value for the entropy in the center of X--ray halos.  Such a
minimum entropy is most likely generated by non--gravitational
processes, in order to produce the observed break in self--similarity
of the scaling relations of X--ray halos.  Likely candidates are SNe
heating, stellar winds, radiative processes, and nuclear activity.  At
present there is no consensus on the level, the source or the time
evolution of this excess entropy.

In this framework a key question is whether the central entropy is the
residual of the entropy originally present in the pre-collapse,
intergalactic medium, or whether it is generated within the halos
after collapse.  Answering this question will allow a more precise
determination of the energetic budget required to build an entropy
floor, and lead towards an understanding of the nature of the sources
of heating.

In this paper we describe a strategy to investigate the physics of the
heating processes acting in groups and clusters.  We show that the
best way to extract information from the local data is the observation
of the entropy profile at large radii in nearby X--ray halos ($z\simeq
0.1$), both at the upper and lower extremes of the cluster mass scale.
The spatially and spectrally resolved observation of such X--ray
halos provides information on the mechanism of the
heating. We demonstrate how measurements of the size of constant entropy
(adiabatic) cores in clusters and groups can directly constrain heating
models, and the minimum entropy value.

We also consider two specific experiments: the detection of
the shock fronts expected at the virial boundary of rich clusters, and
the detection of the isentropic, low surface--brightness emission
extending to radii larger than the virial ones in low mass clusters
and groups.  Both experiments are designed to measure the entropy in
the low density gas far from the core, taking advantage of the
large collecting power of the present X--ray telescopes (in this case
XMM).  Such observations will be a crucial probe of both the physics
of clusters and the relationship of non--gravitational processes to
the thermodynamics of the intergalactic medium.

\end{abstract}


\newpage

\section{Introduction}

The complex thermodynamic evolution of the hot, X-ray emitting, gas in
clusters of galaxies is at the forefront of current efforts to
understand these largest virialized systems. X-ray observations of
cluster number counts, luminosity functions and temperature
distributions indicate little apparent evolution in clusters back to
redshifts as high as $\sim 0.7$ (e.g., Henry 1997, 2000; Rosati et al.
1998). These results provide one of the strongest challenges to high
density cosmological models in which cluster evolution is expected to
be occuring rapidly at low redshifts.

However, these tests are strongly dependent on the thermodynamic
evolution of the intracluster medium (ICM, see Borgani et al. 1999 and
references therein).  In particular, the X-ray properties of X--ray
halos depend on the entropy profile in the ICM. An ubiquitous minimum
entropy, or entropy floor, in the external pre-infall gas, would break the
self--similar behaviour of purely gravitational models, in agreement
with the X--ray data.  In terms of the global X-ray observables
luminosity ($L$) and temperature ($T$), self-similar models predict $L
\propto T^2$, while $L\propto T^{ \alpha}$, with $\alpha\sim 3$, is
observed (David et al. 1993, Mushotzky \& Scharf 1997, Allen \& Fabian
1998, Arnaud \& Evrard 1998, Markevitch 1998), with evidence for a
further steepening at group scales (Ponman et al. 1996; Helsdon \&
Ponman 2000, Xue \& Wu 2000).  Not only can the inclusion of an
entropy minimum successfully reproduce the observed $L\propto T^3$
relationship, but it can also explain the flat density distribution
observed in the cores of clusters and the low evolution of the
$L$--$T$ relation at high redshifts (see Tozzi \& Norman 2000,
hereafter TN).  Recently an entropy floor has been detected in the
core of groups (Ponman, Cannon \& Navarro 1999, hereafter PCN;
Lloyd--Davies, Ponman \& Cannon 2000) providing direct evidence for
this entropy excess in objects with temperatures between $1$ and $3$
keV. Evidence for a breaking of self--similarity also comes from the
observation of a dramatic change in the chemical and spatial
distribution properties of the gas at the scale of groups, below the
observed temperature of 1 keV (Renzini 1997, 1999; Helsdon \& Ponman
2000).

In hierarchical models of structure formation the local ICM is simply
the high redshift IGM accreted into cluster and group scale potential
wells.  An examination of the equation of state of the IGM based on
observations of the high redshift Ly$\alpha$ forest (Schaye et
al. 1999; Ricotti, Gnedin \& Shull 2000) yields an average entropy
level which is at least an order of magnitude lower than that observed
in the core of low temperature clusters, and that needed to explain
the local properties of X--ray clusters and groups.  Physical
processes which could raise the entropy of the early IGM are SNe
feedback, linked to the history of star formation, or radiative and
mechanical processes driven by quasars (see Valageas \& Silk 1999; Wu,
Fabian \& Nulsen 1999; Menci \& Cavaliere 2000).  It is, however, very
difficult to model such processes {\sl a priori}.

Our approach is instead to start from the properties of the ICM as observed in
groups and clusters, and trace back the fundamental processes that drive the
thermodynamic evolution of the gas.  In order to simplify the possible
scenarios, we consider two forms of entropy injection.  First, the excess
entropy may be interpreted as the fossil residual of an initial entropy floor
imprinted in the external, pre--collapse IGM before the epoch of accretion.  In
this case, the heating occurs when only the small, sub--galactic scales have
collapsed, and the gas is heated at about the background density.  Processes
like starbursts can be very efficient in transporting large amount of energy
out of the host galaxies (Strickland \& Stevens 2000).  In the second scenario,
the entropy is the result of heating following the collapse of the baryons in
group--sized potential wells ($M>10^{13} M_\odot$). In this case the gas is
heated at the average density reached in the virialized halos, which can be two
orders of magnitude larger than the ambient density.  These two scenarios imply
a different energetic budget, since for a given entropy level, a larger energy
per particle is required at higher densities.  We hereafter refer to these two
extreme situations as the {\sl external} and {\sl internal} heating cases,
respectively.

In either scenario the enhanced gas density expected around actively accreting,
massive clusters of galaxies (typically a factor of $10$ with respect to the
background value) could make it detectable in emission in the X--ray band even
at distances larger than the shock radius, depending on its temperature and
density.  This gas has not yet been accreted or shocked, hence its entropy is
indicative of the initial value present in the external IGM.  The entropy level
inside the shock radius will be much higher than the external level, as the
result of strong shock heating driven by the accretion process.  The entropy
profile should then decrease towards the cluster center with a power law which
results from the combination of shock heating and previous or ongoing
non--gravitational heating.

At the low end of the mass scale, an external entropy floor gives rise to a
flat, extended, low surface brightness emission, without any shocked
accretion.  In this case the gas has been accreted adiabatically, and the
entropy remains at its initial value everywhere, even in the inner regions. 
With internal non-gravitational heating the profile can be more complex.  Thus
the investigation of lower luminosity, lower mass systems will be a useful
complement to that of the accreting gas surrounding rich clusters.

There are therefore several potentially observable phenomena which
allow us to probe the entropy histories of clusters: accretion shocks,
external warm gas around clusters, highly extended isentropic emission
in groups, and variations in the interior gas entropy profile for both
clusters and groups.  The combined observations of these features over a
range of mass scales can test the internal versus external scenarios.

In the present work we investigate the predictions of the external
and, to a lesser extent, the internal heating scenarios, and
present observational strategies and feasibilities for studying their
physical consequences.  In \S 2 we discuss entropy based models and
accretion processes for X-ray galaxy clusters. In the {\sl external }
heating scenario (\S 2.1) we demonstrate that the regions around rich
clusters are particularly important.  In the {\sl internal } heating
scenario (\S 2.2) the slope of the inner entropy profile can provide an
indication of internal energy injection which followed the accretion.
In \S 3 we present observational strategies to investigate the
described scenarios.  We consider the detection of accretion shock at
the ICM/IGM interface; the expected properties of the accretion shock
regions are described for a range of entropy levels in the external
gas. We then assess the feasibility of observing such accretion shock
regions in real systems, using simulated XMM (see Dahlem et al. 1999)
observations scaled to a nearby cluster (Abell 2029). We consider also
strategies and feasibilities for observing groups, where an
insentropic distribution of gas is expected.  In \S 4 we briefly
discuss stellar processes as a source of entropy, showing the impact
of the proposed observations on the study of the nature of the heating
sources. In \S 5 we present our conclusions.

\section{Entropy-based models of X-ray clusters and groups}

The evolution of the ICM is governed by both dynamics (and the
underlying cosmology) and gas thermodynamics.  A complete treatment of
the physics of the gas necessarily includes shock heating and
adiabatic compression (see the 1D models of Bertschinger 1985, Knight
\& Ponman 1997, Takizawa \& Mineshige 1998, and the 3D numerical
simulations of Evrard 1990, Roettiger et al. 1993, Metzler \& Evrard
1994, Bryan \& Norman 1998), and radiative cooling (see Lewis et
al. 1999).  An expanding accretion shock at the interface of the inner
virialized gas with a cooler, adiabatically--compressed, external
medium, located approximately at the virial radius of the cluster, is
a longstanding prediction from such gravitationally--driven models.
However, as discussed in the Introduction, gravitationally--driven
models predict X-ray properties which scale self-similarly with mass
and fail to reproduce X-ray observations of clusters.

The presence of a minimum entropy in the pre-collapse IGM has been
advocated for some time as a way to naturally break the purely
self-similar behaviour (Kaiser 1991, Evrard \& Henry 1991).  More
recently, a minimum entropy has been detected (PCN) interior to
clusters and all of its consequences have been re--visited.  Models
based on minimum entropy are able to explain the detailed shape of the
$L$--$T$ relation, predict its evolution, and help in explaining the
cores and temperature profiles observed in clusters.  In particular,
we will use the model presented in TN that although having the
limitation of being one--dimensional, does allow
a semi--analytic treatment of shock heating, adiabatic compression and
radiative cooling. With all of these processes being modulated by the
cosmology and  dark matter properties.  The only free parameter in
the model is, in the case of external heating, the initial entropy
value.  In the case of internal heating, the free parameter space
is necessarily larger, depending on the epoch and distribution of the
heating sources within the halo.  Here we will limit the study of the
internal scenario to a simple reference case, outlining only the most
prominent features, without giving an exhaustive investigation of the
many possible heating models.

In the following we describe the two scenarios, referring the reader to the
work of TN and discuss in greater detail the resulting entropy profiles.

\subsection{External heating}

In the external heating scenario, a non--negligible initial entropy in
the IGM introduces a mass scale where strong accretion shocks no
longer form.  Below this mass (at the scale of groups) is an
effectively adiabatic regime, where gas is just compressed into the
potential wells at constant entropy.  The observed $L\propto T^3$
relationship is essentially produced by the resulting flattening of
the density distribution in cluster cores, when shocks turn off
completely (see Balogh, Babul \& Patton 1999, Tozzi \& Norman
1999 (TN)).

After the accretion, the entropy of each accreted shell of gas is kept
constant, as long as the density is low and the cooling time (defined as
$t_{cool}\propto \rho^{-1} T^{1/2}$ for $kT>2$ keV) is correspondingly large. 
Cooling can be important in the central regions, depending also on the initial
entropy level.  Indeed, if the initial entropy is too low, cooling is the
dominant process, leading to an excessive amount of cooled baryons.  In this
strong cooling regime (actually disfavoured by many observations and not
considered further here), the semianalytical model breaks down.  As stated
above, further (internal) non--gravitational heating processes are not included
in our external scenario.

The mass scale of shock formation is governed by the value of the IGM entropy
$S \propto {\rm log}(K)$, where $K\equiv kT/\mu m_H \rho^{2/3}$ (here we assume
that $\mu =0.59$ for a primordial IGM). The value of $K=K_*$ that produces a
good fit to the local $L$--$T$ relation is $K_*=(0.2\pm 0.1) \times 10^{34}$
erg cm$^2$ g$^{-5/3}$ (see TN), which needs to be in place by at least the
turn--around epoch for each shell.  For a given entropy level, the temperature
of the gas at the universal background density is derived as $kT\simeq
3.2\times 10^{-2} K_{34} \times 10^{34}(1+z)^2$ keV (assuming the standard
nucleosynthesis baryon density value).  In terms of energetic budget, this
value corresponds to a lower limit of $kT_{min} \simeq 0.1 (K_{34}/0.4)$ keV
per particle in a $\Lambda$CDM universe ($kT_{min} \simeq 0.04\, (K_{34}/0.4)$
keV in a tilted CDM with $\Omega_0=1$). Note that this value is within the
energy budget expected from SNe heating (see Loewenstein 2000).

The observed entropy floor in the core of groups is about $K_{cl}\simeq 0.08\,
h^{-1/3} \times 10^{34}$ erg cm$^2$ g$^{-5/3}$ (corresponding to $S\equiv
kT/n_e^{2/3}\simeq 80\, h^{-1/3}$ keV cm$^2$ using the definition of PCN,
within an uncertainty of a factor of 2; see PCN, Lloyd--Davies, Ponman \&
Cannon 2000).  We stress that in small mass halos, the central entropy excess
is always expected to be smaller but close to the initial value at the epoch of
accretion since the accretion proceeded adiabatically, and the cooling has been
inhibited by the initial entropy level itself.  Possibly, the observed value of
$K_{cl}$ may be different from that in the external IGM being accreted at $z=0$
(the one we would want to detect) if substantial evolution in the entropy
occurred since the accretion epoch.  Indeed, since groups form typically at
early lookback times, their central entropy could actually be lower than the
present--day level in the external IGM if $K_*$ grows with the cosmic epoch. 
We will consider this when discussing possible scenarios for actual
observations.

Interestingly, the simplest assumption $K_*\geq K_{cl} = constant$ actually
reproduces the observed break in scale invariance rather well.  At low masses
($10^{13} - 10^{14} h^{-1} M_\odot$) the emitting gas will be very extended,
without any discontinuity in the entropy since there are no shocks which would
otherwise separate the accreting IGM from the accreted ICM. Indeed, data
suggest that the total luminosity of a group can vary by a factor of 2 or 3
after the inclusion of the undetected, low surface brightness emission
extending up to the virial radius (see Helsdon \& Ponman 2000).  This is
consistent with the prediction from the analytic model in TN, where, for a
given temperature, the total luminosity (i.e., including the emission from all
the gas within the shock radius) can exceed by a factor larger than 3 the
luminosity included within $100 h^{-1} $ kpc, a radius often used to define the
luminosity of poor groups.  As we will discuss in \S 3, detecting the missed
emission from loose groups is one of the scientific aims of our proposed
observational scheme.

Moving to higher mass scales, the initial entropy level has less effect on the
outer cluster regions.  The average entropy is dominated by the entropy
produced in strong shocks. However, the initial entropy still contributes to
building a central density core, even if the effect of cooling starts to be
important in eroding the entropy plateau.  An important point is that, in
contrast to low mass systems, the gas density is still significant at radii
larger than the virial radius.  Specifically, at distances larger than the
shock radius itself, the overdensity of the accreting gas will be $\simeq 10$
with respect to the background density.  The entropy level of this gas is by
definition the value $K_*$.  We expect a considerable amount of such diffuse
gas around massive halos.  In particular, the most massive clusters are likely
to be still accreting significant amounts of matter in most cosmologies, since
they are the last objects to form in any hierarchical universe. The total
accretion rate of matter for a cluster of $8$ keV (roughly corresponding to a
mass of $\simeq 10^{15}\, h^{-1}\, M_\odot$) is expected to be on average quite
high at $z=0$.  The predicted average mass growth in baryons, computed in the
extended PS framework (see, e.g., Lacey \& Cole 1993), is about $f_b \,
0.24\,\times 10^{15} M_\odot$/Gyr for $\Omega_0=1$ with a tilted, cluster
normalized CDM spectrum (tCDM) and $f_b\, 0.08\,\times 10^{15} M_\odot$/Gyr for
$\Omega_0=0.3$ and $\Lambda = 0.7$ ($\Lambda$CDM).  Here $f_b$ is the universal
baryonic fraction.  In the following we will use the standard nucleosynthesis
value $f_B= 0.02 h^{-2}/\Omega_0$ for $\Lambda$CDM.  This value gives similar
baryonic accretion rates in $\Lambda$CDM and tCDM.  However, we are forced to
use a value at least two times larger in tCDM, in order to have the average
baryonic fraction in halos as high as $\simeq 15$ \%, as observed (White et al.
1993).  As a general feature, the accretion rates are correspondingly higher at
higher redshifts for the same masses.  However, in this current work we will
focus on $z\simeq 0$.

The break scale between adiabatic (low masses) and shock (large
masees) regimes can be investigated by studying the dependence of the
infall velocity $v_i$ of the accreting IGM on the total mass of the
system.  Approximating the infall as an adiabatic flow, we calculate
the infall velocity at the shock radius, where the gas is expected to
be shocked and reach hydrostatic equilibrium.  The effect of a
constant entropy minimum $K_*$ on the infalling IGM is to introduce a
compression term proportional to the square of the sound speed
$c_s^2=\gamma K_*\rho_e^{2/3}$, where $\rho_e$ is the external
baryonic density and $\gamma = 5/3$ is the adiabatic index for a
monoatomic gas.  In fact, part of the gravitational energy goes into
compression, to give for the infall velocity:
\begin{equation} 
{{ v_i^2}\over 2} = {{ v_{ff}^2}\over 2} +\Delta W - {{ c_{s}^2}\over
{\gamma -1}} + {{ c_{s}^2}\over {\gamma -1}} \Big({{ \rho_{ta}}\over
{\rho_e}}\Big)^{\gamma -1}\, , \label{vi}
\end{equation}
where $v_{ff}$ is the free fall velocity, and $\Delta W$ is the
contribution added to $v_{ff}^2/2$ to have the total work done by the
gravitational potential (see TN).  The fourth term on the right hand
side results from the initial condition $v_i=0$ for a gas shell at the
turnaround radius, when the gas had a density $\rho_{ta}\simeq
\rho_{back}$.  The compression term carries an increasing fraction of
the total gravitational energy when the system mass is lower, or,
since the sound speed is proportional to $K_*^{1/2}$, when the entropy
is higher.

\begin{figure}
\centerline{\psfig{figure=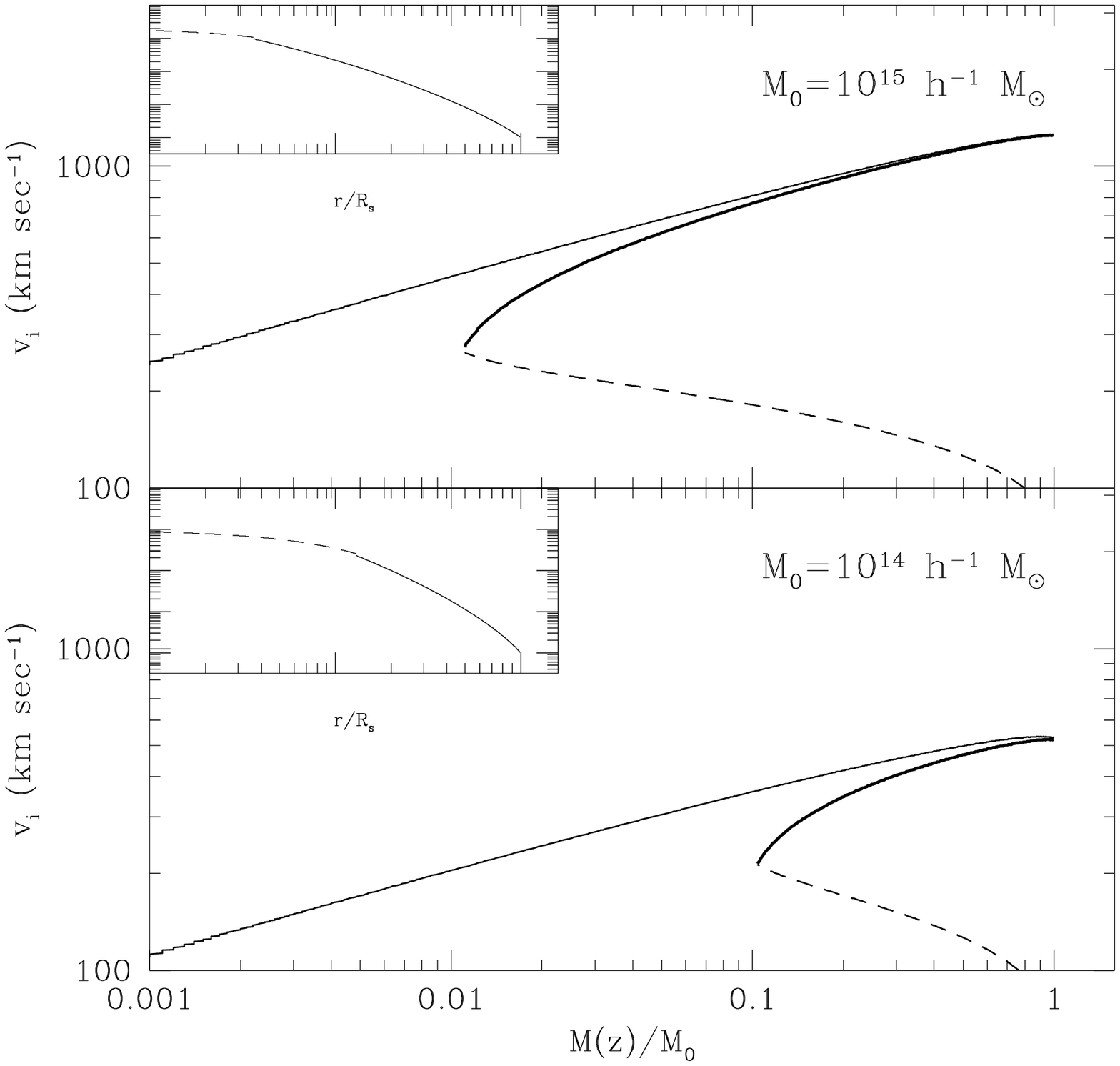,height=5.in}}
\caption{ The infall velocity at the shock radius of each accreted gas
shell as a function of the virialized mass. Here we assumed $K=
0.3\times 10^{34}$ erg cm$^{2}$ g$^{-5/3}$ in a low density
($\Omega_0=0.3$) flat cosmology, for a final mass at $z=0$ of
$10^{15}\, h^{-1} M_\odot$ (upper panel) and $10^{14}\, h^{-1}
M_\odot$ (lower panel).  The straight line is the free fall velocity
$v_i\simeq M^{1/3}$, while the thick line is the infall velocity of
the baryons, and the dashed line is the sound speed $c_s$.  When
$v_i<c_s$ the accretion process is entirely adiabatic, and the
accreted gas forms an adiabatic core (dashed line) in the
 density profiles shown in the inset boxes (for simplicity
 cooling is not included in the calculation of these profiles). 
\label{fig1}}
\end{figure}

The infall velocity can then be compared with the sound speed in the
infalling gas, to test whether $v_i>c_s$ and shocks can develop.  In
Figure \ref{fig1} the infall velocity computed at the shock radius is
plotted as a function of the virialized mass, which is in turn a
function of the redshift (here we have assumed an average mass growth
for the dark halo commensurate with a $\Lambda$CDM universe).  In the
above picture only the external entropy level is needed to determine
the transition between the adiabatic and the shock regime. 
The external density $\rho_e$, which determines the sound speed in the
IGM, is obtained by imposing mass conservation at the accretion
radius, plus the assumption that the accreted baryons are a constant
fraction of the total virialized mass (see TN).

At early epochs, when the virialized mass is still low, the compression term is
important and the infall velocity is lower than the sound speed.  In this case,
the accretion of the IGM proceeds entirely adiabatically, giving rise to an
adiabatic core (insets Figure \ref{fig1}.  As  the virialized mass grows, the
infall velocity eventually becomes larger than the sound speed, marking the
epoch when shocks dominate (here we have neglected the small velocity of the
shock front in the cluster rest frame).  The infall velocity then
asymptotically approaches the free fall velocity of the system.  In Figure 1 an
external constant entropy of $K_{34} = 0.3$ (where $K_{34}$ is in units of $
10^{34}$ erg g$^{-5/3}$ cm$^2$) has been assumed for a low density
($\Omega_0=0.3$) flat cosmology, for objects of mass $10^{14} h^{-1} M_\odot$
and $10^{15} h^{-1} M_\odot$.  At lower masses, the transition from the
adiabatic to the shock regime occurs later, giving rise to a relatively larger
adiabatic core.  From Figure \ref{fig1} we can also see that for low mass
systems ($M<10^{14}M_\odot$), a growing fraction of the accreted baryons
retains the pre-collapse entropy level. This fraction approaches unity at the
scale of poor systems and groups, providing self-consistency with our
expectations of isentropic gas in groups.  Note that the transition beteween
the adiabatic and the shock regime is marked by a transition radius $r_t$
within which an approximately constant baryonic mass in contained.  This can
provide a meaningful observable as further discussed in \S 3.

As a consequence of the above picture, the entropy profiles change dramatically
along the mass sequence: flat in low mass halos, steep and discontinuous for
large masses.  The steep part of the profile corresponds to strongly shocked
gas, while the flat part is the adiabatically accreted gas.  External to the
shock, the accreting gas is simply adiabatically compressed.  In the following
sections we will describe in greater detail the resulting entropy profiles.

\subsection{The entropy profile with external heating}

We focus first on large mass scales, where a strong accretion shock
is expected irrespective of the initial entropy level. Such an accretion
shock is likely to occur at approximately the virial radius, where the
gas density can typically be a factor $\simeq 1000$ lower with respect
to that at the cluster center.  A simple relation exists between the
density jump and the temperatures of the hot internal, and colder
external, gas (Landau \& Lifshitz 1959, see Cavaliere, Menci \& Tozzi
1997, 1999):

\begin{equation}
\rho_i/\rho_e = 2 \Big(1-{{T_e}/{T_i}}\Big)+ \sqrt{4
\Big(1-{{T_e}/{T_i}}\Big)^2+{{T_e}/{T_i}}}\, , 
\label{g} 
\end{equation} 
where $\rho_i$ and $T_i$ are the internal gas density and
temperature. The external density $\rho_e$ and temperature $T_e$ refer
to the infalling gas just prior to being shocked.  Note that $T_e$ is
{\em not} simply the temperature of the field IGM. The accreted IGM
will experience adiabatic compression prior to reaching the accretion
shock, thus $T_e = \mu m_p K_* \rho_e^{2/3}$.  The overdensity of the
baryons with respect to the background value is expected to be $\simeq
10$ for rich clusters both in $\Lambda$CDM and in tCDM.  This would
correspond to temperatures of $kT_e \simeq 3.2 \times 10^{-2} K_{34}
\, \delta^{2/3}\simeq 0.15\, K_{34}$ keV at $z=0$.  However, since in
tCDM we are forced to assume a baryonic density larger than a factor
of $2$ (with respect to the standard nucleosynthesis value), the
external temperature $T_e$ for a given $K_*$ and $\delta$ will be
about $60$ \% larger.  

To compute the internal density at the shock boundary, we could use
the detailed and self--consistent density and temperature profile of
the ICM resulting from a minimum entropy model, as derived in TN.
However, the density profiles can be fitted to a good approximation
with a $\beta$ model (Cavaliere \& Fusco Femiano 1976), at least at
large radii where the cooling time is large, as it is shown in Figure
\ref{fig2}.  Therefore for simplicity we use it to approximate the gas
density internal to the shock:
\begin{equation}
\rho = \rho_c\Big(1+(r/r_c)^2\Big)^{-{3\over 2}\beta}\, ,
\label{beta}
\end{equation}
where $r_c$ is the core radius.  For a flat temperature profile, the
observed X-ray surface brightness at the shock radius $r_s$ can be
written as:
\begin{equation}
\Sigma = \Sigma_c {\Big(1+(r_s/r_c)^2\Big)^{-3\beta+1/2 } }\, .
\label{sb1}
\end{equation}
In this case a simple inversion from surface brightness to density
within the shock radius is given by:
\begin{equation}
{{\rho}\over{\rho_c}} = \Big( {{\Sigma} \over{\Sigma_c}}\Big)
^{1/(2-1/3\beta)}\, .
\label{sb2}
\end{equation}
For a $\beta$ model, the discontinuity in the surface brightness
expected at the shock is approximately
$(\rho_i/\rho_e)^{2-1/3\beta}(T_i/T_e)^{1/2}$ (for $T_e\geq 1$ keV)
with respect to the extension of the pure beta model to the shock
radius. In the case in which there is no shock ($\rho_i/\rho_e=1$) and
the temperature profile decreases adiabatically as $T\propto
\rho^{2/3}$ (expected in small groups), we can use the same functional
form, replacing $\beta$ with an effective $\beta'$ which accounts for
the mild dependence of the emissivity $\epsilon$ on temperature.  In
the case of pure bremsstrahlung and temperature $kT>2$ keV, $\epsilon
\propto T^{1/2}$ and $\beta'={7\over 6} \beta$ (see Ettori 1999).  At
lower temperatures we must include the contribution from line
emission, which is significant in the wide energy band of XMM
($0.1-12$ keV).  For $0.1<kT<2$, $\epsilon \propto constant$ with good
approximation if the metallicity is about one third solar, virtually
removing the temperature dependence and giving $\beta'=\beta$ again.
The entropy jump can therefore be detected when both the X-ray surface
brightness (from which density is determined) and the temperature are
measured at the shock radius.  In Figure \ref{fig2} the surface
brightness and the emission weighted temperature profiles are shown
for three relevant cases of the external scenario, the same that will
be discussed in detail in \S 3.

The effect of the accretion shocks is to raise the entropy over its
initial (external) level.  Since the transition from the adiabatic
accretion to the shocked accretion is very fast (with the shock radius
rapidly approaching the virial one, see Figure \ref{fig4}, third
panel), the transition between the two regimes is recorded in the
entropy profile as a sudden change of slope at the transition radius
$r_t$.  In the inner part the adiabatic core is visible (though
eventually it will be partially erased by cooling), while in the
outer, shocked regions a featureless power law profile is expected.  A
reference slope for this power law can be derived to a first
approximation with the usual assumption of an isothermal profile:
$\rho\propto r^{-2}$ gives $K\propto r^{4/3}$.  This value is close to
the $K\propto r^{1.1}$ predicted from the model (see TN), where a
temperature gradient is present due to further adiabatic compression
after accretion, and the density distribution is somewhat steeper than
$-2$.  In other words, the gas in cluster is well described by a
polytropic distribution with a polytropic index $\gamma_p\simeq 1.2$
(see Loewenstein 2000; TN).  The expected entropy profiles in the
external heating scenario are shown in Figure \ref{fig3} for two
objects of $M=10^{15}H^{-1} M_\odot$ and $M=10^{14}H^{-1} M_\odot$.
We can see the entropy core partially erased by cooling within $r<0.1
\, R_{vir}$, and the shocked gas with the characteristic slope
$K\propto r^{1.1}$ at $r\simeq 0.1-0.3 h^{-1}$ Mpc.

\subsection{Internal heating}

How do these predictions change in the internal heating scenario?  A
first major difference is that the gas is heated when it is at much
higher densities, and the change of entropy for a given energy input
is consequently lower.  In other words, to reproduce the breaking of
the scale invariance in X--ray halos, a larger energy budget is needed
in the internal scenario with respect to the external one. In the
internal heating case, the IGM is essentially cold when it is
accreted, and the gas always experiences strong shocks, even in low
mass objects.  In this case, the gas may be
detected via emission in the UV band (and may be related to the UV
excess detected around nearby clusters, see Lieu, Bonamente \& Mittaz
2000).  Alternatively, such gas may be seen in absorption against
bright background sources, in the X--ray band if $kT\simeq 0.1 $ keV
(see Hellstein, Gnedin \& Miralda--Escude' 1998), or in the UV band if
$kT\simeq 0.01 $ keV.  In the last case OVI, which peaks at $\simeq
0.03 $ keV in collisional equilibrium, is the best diagnostic
(K. Sembach, private communication).

It is worth recalling that in the absence of any heating, in the
central regions of halos the entropy gained by shock heating alone
is not enough to prevent the gas from cooling.  In fact, the
absence of an initial extra entropy would result in a cooling
catastrophe (see White \& Rees 1978, Blanchard, Vall Gabaud \& Mamon
1996).  Furthermore, the combination of shock heating and central cooling
(without
additional heating) is, in fact, not able to generate an entropy floor
by the selective removal of the lowest entropy gas in the very center
(see TN), a mechanism sometimes advocated to explain the entropy
plateau (see PCN).  

In the internal heating scenario, the number of free parameters is
larger than one and dependent on the model used to describe the heating
sources.  Here we assume a simple phenomenological model, with a
distribution of sources of equal mass (or output) given by a King profile
with a
large core (about $1/2$ of the virial radius).  The number of the
heating sources is then normalized to the total mass of
the given halo.  The absolute number density of the sources is clearly
degenerate with the average heating rate associated with each source,
since only the global heating (as a function of the radius $r$) is
relevant to the final entropy profile.  Therefore, for each heating
model we quote only the average heating per particle released up
to the present epoch.  Not that the heating is defined as the
amount of energy dumped into the ICM, which can clearly be different from
the total
energy budget of the sources, depending on the gas heating efficiency.

We note that the difference in the energy budget between the internal
and the external heating, is further exacerbated by cooling.  Indeed,
if the internal heating is not rapid enough to keep the density low and
prevent
further cooling, the densities in cluster centers will be always very
large, and the same final entropy profile will require a very large
energy budget.  A direct consequence is that if the heating is not
large enough, the energy input will be rapidly re-emitted by the high
density gas, no matter how much energy has been released (see, e.g.,
Lewis et al. 1999).

The heating rate must have a dependence on $z$; indeed, if the
heating starts in the halo but at very early epochs, before a
non--negligible amount of mass of the final halos has been accreted,
the gas never reaches high overdensities and the
properties of the external scenario are reproduced.  Here we assume
(motivated by the need to reproduce the observed $L$-$T$ relation) that
the
heating rate peaks at $z\simeq 1$, with an exponential decline at higher
redshifts, and a mild power law decline $\propto (1+z)^2$ for $z<1$.
The final energy budget is, in this case, mass dependent, since the
number of sources is larger in higher mass halos.  Moreover, enhanced
heating is expected in the center, where the density of the heating
sources is higher.

For the assumed peak redshift $z=1$, we calibrate the energy budget by
requiring consistency with local properties (e.g., fitting the
$L$--$T$ relation).  We find that a total budget of about $1-2$ keV
per particle in clusters ($M=10^{15} h^{-1} M_\odot$) and $0.5-1$ keV
in small clusters ($M=10^{14} h^{-1} M_\odot$) can reproduce
approximately the scaling relation for X--ray halos (see also Figure
\ref{fig4}).  Despite the uncertainties in the internal heating model,
we always find that the energetic budget is more than an order of
magnitude larger with respect to the external scenario in order to
reproduce approximately the $L$--$T$ relation.  This estimate is
robust and it is expected also on the basis of a simple analytical
calculation, since the typical overdensities within virialized halos
are of the order of few hundred, while the typical overdensities in
the external accreting gas are $\delta \leq 10$ (see TN).  The
energetic budget in the internal scenario may be too high to be
provided by SNe heating only.  In this perspective, the comparison of
the external and internal scenario may put constraint on the nature of
the heating sources.

In Figure \ref{fig3} we show the resulting entropy profiles, in comparison
to those of the external scenario.  In a massive halo, the relatively flat
distribution of heating sources results in a flatter entropy profile
in the center $K\propto r^{0.5}$, while in the external regions the
same profile of the external case $K\propto r^{1.1}$ is recovered.
The slope of the inner profile depend on the total amount of energy
injected, as shown by the labels in Figure \ref{fig3}.  The same
amount of heating in a smaller halo ($M\simeq 10^{14} h^{-1} M_\odot$)
results in a larger entropy core that emerges in the central regions.
In the regions where a negative entropy gradient developes, we expect
instabilities and mixing.  We do not show the case of very small
halos, since large negative entropy gradients develop with consequent
instabilities that cannot be included in the present treatment.  The
effect may be real, in the sense that in small halos the effect of the
internal heating may disrupt the profile resulting from
adiabatic/shocked accretion, and can eject virtually all the gas from
the potential well, giving a patchy and irregular surface brightness.
This possiblity should be investigated with fully 3D numerical
simulations.

In principle, measuring the entropy profile at a radius $\simeq 0.1
R_{vir}$ in high and medium mass halos, can reveal a signature of the
internal heating scenario as a departure from the profile
expected in the external scenario. However, we emphasise that this is just
an example based on a particular choice of the internal heating
distribution, and in some cases the internal heating may result in an
entropy profile very similar to that of the external scenario.  A more
comprehensive investigation of the parameter space in the case of
internal heating will be presented in a future work.

It is worth noting, nevertheless, that observations at a radius
$r\simeq 0.1 R_{vir}$ will be less difficult with respect to those
at the shock radius, due to the higher surface brightness.  The signal
will be much higher and the entropy profile can be reconstructed in
greater
detail (see PCN).  In principle, it will also be possible to produce 
entropy maps inside clusters.  In the assumption that the ICM has not
been stirred due to massive merger events (an assumption which is
implicit for the spherical model used here and described in TN), the
entropy maps would trace the patches of major heating internal to the
cluster.

In Figure \ref{fig4} we show the time evolution of a cluster, a small
cluster and a group in a $\Lambda$CDM cosmology for the internal
(dashed lines) and external heating (continuous line) scenarios.  The
external scenario assumes $K_{34} = 0.3$, while the internal scenario
assumes an energy budget of $\simeq 0.9$ keV per particle.  The
lowest mass ($M=10^{13} h^{-1} M_\odot$) is not shown for the latter
model.  At $z=0$ the
total luminosities and the emission weighted temperatures are quite
similar, and it is not possible to distinguish the two scenarios from
the statistical properties of the X--ray halo population only; both
scenarios fit the $L$--$T$ relation (as can be confirmed from
the final luminosities and temperature at the different scales; see
however TN).  The shock radius is much closer to the virial one in the
external scenario, since the external gas is cold and its infall is not
slowed
by the pressure support while it is accreted.  The largest difference
in the shock position are predicted at low masses, where,
unfortunately, the shock feature is currently hard to detect due to the
low surface brightness.

Another way to break the degeneracies between the internal
and the external scenarios, is to look at the global properties of
high redshift halos, for which the expected differences are
larger.  Indeed,
the entropy level is the major driver of the evolution of the global
properties of X--ray halos (see also Bower 1997).  At large redshifts
($z\geq 1$) the luminosity and temperature evolution depends on the
intensity and the timescale of the non--gravitational heating.  In the
internal case, the luminosity and temperature have a flatter time
dependence.  The shock radius is always close to the virial radius
(third panel) in the internal heating scenario, beacause the entropy
of the external gas is always negligibly small.  Thus the epoch and
distribution of the heating affects global quantities, such as the
contribution
of
X--ray halos to the X--ray background (see Wu, Fabian \& Nulsen 1999,
where the gas is heated along the merger tree of  halos and
requires an average extra energy of $1-3$ keV per particle).  However,
we recall again that the many parameters produce a large degeneracy in
the internal scenario.  The external scenario, instead, provides  better
defined predictions, since it depends only on the initial value of the
entropy.  This further strengthens our claim that the best way
to probe the thermodynamic history of the ICM is by looking at the
entropy profile of nearby halos rather than the global properties
of unresolved distant halos.

\section{Simulated Observations and Feasibility}

Here we focus on the observation of nearby halos.  This is the
strategy that we propose as the best way to investigate the
thermodynamic history of the baryons, taking advantage of the spatial
and spectral resolution of present day X--ray missions.  In particular
the external heating scenario can be tested by the detection of the
external entropy level around present day clusters, allowing the
accretion shock itself to be located.  A power law entropy profile,
$K\propto r^{1.1}$, is expected between the shock radius and the
transition radius $r_t$.  Within this last radius, a flatter entropy
profile will mark the original entropy plateau, partially eroded by
cooling.

As described in \S 2, the two dominant, gravitationally--driven,
mechanisms for changing the thermodynamic state of cluster gas are
shock heating and adiabatic compression.  While shock heating occurs
principally at the accretion radius, adiabatic compression will occur
both interior and exterior to this radius. As shown above, adiabatic
compression of gas during accretion (prior to being shocked) will
raise the external gas temperature to values dependent on the initial
entropy. The average entropy in the external IGM at $z=0$ can span an
order of magnitude and still give a good fit the $L$--$T$ relation.

In the case of a constant entropy, the range is $K_* = 0.2 - 0.4
\times10^{34}$ erg g$^{-5/3}$ cm$^2$, which corresponds approximately
to pre--shock (adiabatically--raised) temperatures of $kT_e\sim 0.1$
keV maximum.  Another interesting possibility is to assume a strong
evolution in the entropy, of the form $K_*\propto (1+z)^{-2}$.  This
case gives a good fit to the local $L$--$T$, and at the same time has
a value as high as $K_*(0)=3 \times10^{34}$ erg g$^{-5/3}$ cm$^2$ in
the external gas,corresponding to a temperature of $\simeq 1 $ keV.
Despite the very high final value of the entropy, the total (average)
energetic budget is less than $ 0.1$ keV per particle, even lower than
in the case with constant $K_*\simeq 0.3$.  The reason is that in the
constant entropy case, most of the energy is released at high
redshift, when the density is higher, while if $K_*\propto
(1+z)^{-2}$, most of the energetic budget is released only at small
$z$.  We also consider this case since from the observational point of
view it is one of the most tractable; an external temperature of
$\simeq 1$ keV makes the gas (at an overdensity of $\simeq 10$)
detectable in emission.  We note that such high temperatures in the
infalling gas can  also be achieved in the case in which the gas is
previously gravitationally shocked in filaments (see Cen \& Ostriker
1999 and references therein).  As we will discuss later, this
gravitational contribution to the external entropy may help in attaining
a large value of $K_*$ in the outskirts of clusters.

Once the external level of the entropy is assumed, the other relevant
piece of information is that the gas density immediately exterior to
the shock will be no more than a factor of $\sim 4$ lower than that at
the inner shock boundary, following equation \ref{g}.

\begin{figure}
\centerline{\psfig{figure=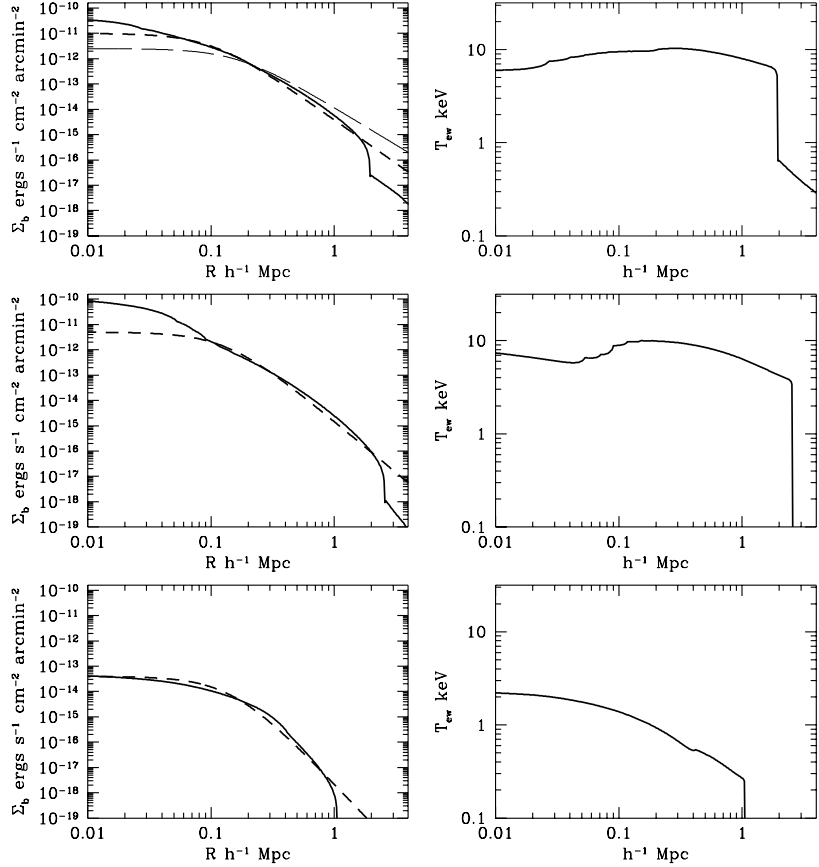,height=5.in}}
\caption{Surface brightness and projected temperature profiles
extended out of the shock radius for three relevant cases in the
external scenario (the entropy and cosmologies have been chosen to emulate
the required properties).  
 Top panels: $M=1.4 \, \, h^{-1} 10^{15} M_\odot$,
tCDM, $K_*=3 \, \, (1+z)^{-2}$; middle panels: $M=1.4 h^{-1} 10^{15}
M_\odot$, $\Lambda$CDM, $K_*=0.3$; lower panels: $M=5 \, \, h^{-1}
10^{13} M_\odot$, $\Lambda$CDM, $K_*=0.3$.  The corresponding fits
with a beta model are shown with a short dashed line.  In the first
panel, the long dashed line shows the beta profile used in the
simulated observations (\S 3).  In this calculation of the predicted
surface
brightness the cooling is included (see TN).
\label{fig2}}
\end{figure}

\begin{figure}
\centerline{\psfig{figure=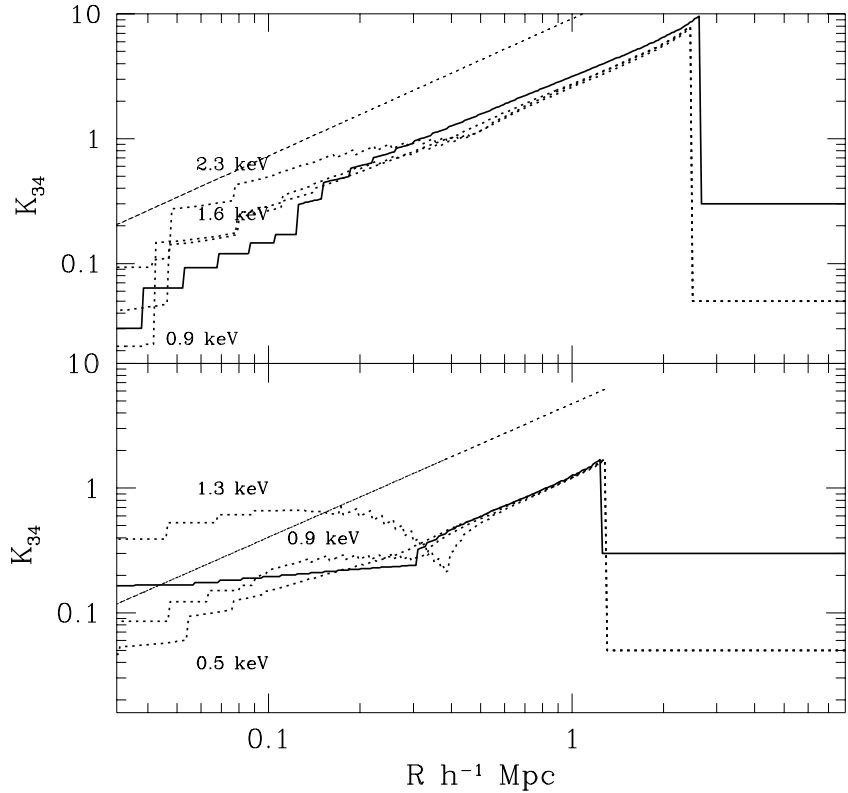,height=5.in}}
\caption{The entropy profiles in the external (continuous lines) and
the internal (dotted lines) scenario for a rich cluster ($M=10^{15}
h^{-1} M_\odot$, upper panel) and a small cluster ($M=10^{14} h^{-1}
M_\odot$, lower panel).  The entropy level is $K_{34} = 0.3$
in the external scenario.  The internal scenario a total budget of
$0.5-2$ keV is released, as shown by the labels.  The dashed line is
the reference power law of $1.1$.
\label{fig3}}
\end{figure}

\begin{figure}
\centerline{\psfig{figure=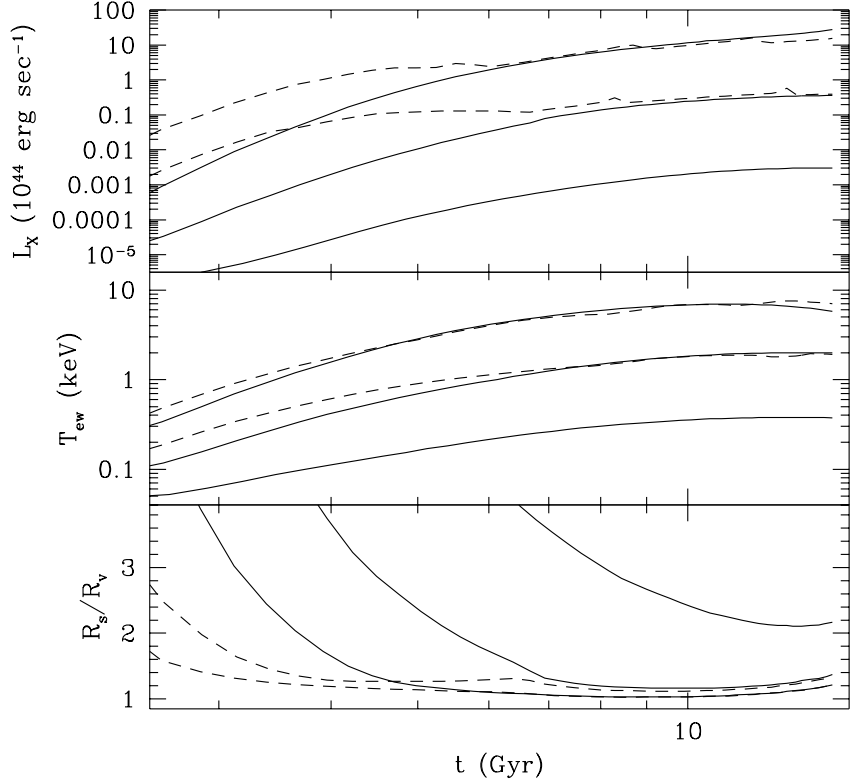,height=5.in}}
\caption{Evolution of luminosity, temperature and shock radius for
halos of $M=10^{15}-10^{14}-10^{13} h^{-1} M_\odot$ (from top to
bottom).  The external scenario with $K_{34}=0.3$ is shown as a
continuous line, the internal (with an energy budget of $0.9$ keV)
with dashed lines.  A $\Lambda$CDM cosmology has been adopted.
\label{fig4}}
\end{figure}

It is also a consequence of the increasing shock strength at larger
radii and of the adiabatic compression that a mildy negative (radially
decreasing) temperature gradient is expected, in good agreement with
current observations of clusters (Markevitch 1998).  The temperature
gradients are expected to be stronger when the entropy distribution
gets flatter, until the adiabatic limit $T\propto \rho^{2/3}$.  For
simplicity, in rich clusters we will consider an isothermal
distribution of gas within the shock radius, since the predicted
temperature profiles can be well approximated as constant (the
predicted polytropic index is $\gamma_p \simeq 1.1$, where $\gamma_p =
1$ is the isothermal case, see Figure \ref{fig2}).

What is the best strategy to investigate the described scenarios?
Recent X-ray data lack the necessary combination of both spatial and
spectral resolution to have routinely detected the accretion shock and
detailed entropy profiles of clusters.  The same situation applies for
low luminosity groups (for $kT<1$ keV, the luminosity is often defined
within a fixed radius of $100 \, h^{-1}$ kpc, see Ponman et al. 1996).
ROSAT for example, while its limiting surface brightness was quite low
(a typical background level $\sim 1\times 10^{-15}$ erg s$^{-1}$
cm$^{-2}$ arcmin$^{-2}$ in the 0.5-2 keV band), had a point spread
function width from $\sim 20-60$ arcsec and insufficient spectral
sensitivity to constrain temperatures to the precisions
required. Attempts to push the capabilities of the ASCA X--ray
satellite to their limits and observe this accretion shock in archival
data of nearby clusters failed, mainly due to the poor ASCA
point-spread function (Gendreau \& Scharf, private communication).

Current missions should however be well suited to detecting cluster
accretion shocks and entropy profiles.  Chandra's high spatial
resolution ($\sim 1-10$ arcsec) may allow details of the spatial
structure of a shock region to be investigated.  With an effective
area of $\sim 4600$ cm$^2$ at 1 keV, XMM has approximately 10 times
higher throughput than ROSAT, and combined with a $\sim 6-15$ arcsec
PSF and excellent spectral resolution is ideally suited to this
task. XMM can, for example, detect an accretion shock in the nearby
Perseus cluster ($z=0.018$, $L= 2.8\times 10^{44}$ erg s$^{-1}$ in the
2-10 keV band) with an exposure of the order of 20 ksec. However, in
this case, the area to be searched is extremely large compared to the
field of view of XMM.

For the lower mass groups the detection, or non-detection, of the
accretion shock is much more difficult. However, determining the
emission profile beyond the regime currently studied ($\sim 100 \,
h^{-1}$kpc) and the nature of the entropy profile will be quite
feasible.  In both cases (rich and poor systems) a key observational
criteria will be the ability to detect emission at a level of $\sim
10^{-16}$ erg s$^{-1}$ cm$^{-2}$ arcmin$^{-2}$ (see Figure
2). Additionally, the ability to accumulate sufficient counts to
constrain gas temperatures to precisions of 10-20\% will be necessary.
In general, with a typical background, we find that to measure the gas
temperature with a precision of $\sim 10-20$\% requires at least $\sim
1000-2000$ source photons respectively for low ($\sim 1$ keV) and high
($\sim 8$ keV) temperatures. This is a consequence of lower
temperature spectra having more photons on the exponential cutoff,
where the impact of temperature is strongest.

As an observational baseline for rich clusters we have chosen the
Abell 2029 system. At a redshift of $z=0.0767$ and with $L_{2-10 keV}
= 2.07 \times 10^{45}$ erg s$^{-1}$ ($h=0.5$) and $kT=7.8$ keV (David
et al. 1993) this cluster presents an optimal angular scale ($\sim 30$
arcmin) and surface brightness. In addition Abell 2029 is a strong
cooling flow cluster, and, at least in the inner regions, appears to
be in equilibrium with no sign of merging of cluster subunits (Sarazin
et al 1998). The core radius of 0.164h$^{-1}$Mpc, corresponds to $\sim
2.5$ arcmin. Assuming $\beta=2/3$, we obtain an estimated surface
brightness at the shock of $\sim 1 \times 10^{-16}$ erg s$^{-1}$
cm$^{-2}$ arcmin$^{-2}$, in close agreement with our model (see Figure
2).

We can now ask what would be required to measure the entropy profile
to an emission level of $\sim 10^{-16}$ erg s$^{-1}$ cm$^{-2}$
arcmin$^{-2}$, i.e., to the shock radius.  Combining the expected
count rates (from XSPEC) for the PN +2MOS for a 7.8 keV plasma and the
expected background counts we estimate that to achieve an emission
detection of $\sim 6\sigma$, and at least 2000 cluster photons,
requires $\sim 70$ ksec and counts accumulated from $\sim
270$ arcmin$^{-2}$.  Given the angular dimensions of A2029 we could
meet these criteria with 4 XMM pointings of 70 ksec, equally spaced
around the expected $\sim 20$ arcmin shock radius.  The detection of
the accreting gas beyond the shock radius is more difficult, since the
surface brightness of the external gas can be an order of magnitude
lower.  We therefore perform a more realistic simulation in order to
assess the feasibility of its detection.

Using QUICKSIM (Snowden 1998) and XSPEC we have simulated a range of
XMM observations of cluster gas emission. The simulated cluster
(group) is orientated such that the XMM field of view is centred on
the shock radius.  We simulate both the PN and two MOS EPIC
cameras. The internal and cosmic background count rates in the
$0.1-12$ keV band at the coordinates of Abell 2029 are estimated to be
$3.67\times 10^{-3}$ ct s$^{-1}$ arcmin$^{-2}$ (PN) and $1.11\times
10^{-3}$ (MOS) and are included.

We began by simulating a 200 ksec XMM observation of A2029 with a
pointing offset $\sim 20$ arcmin from the cluster center (i.e. still
within the expected shock radius).  The external entropy level is set
to $K_{34} = 3\, (1+z)^{-2}$.  The assumed surface brightness profile
is the one shown in Figure \ref{fig2} as a long--dashed line, which is
flatter than the average expectation but better resembles the
realistic case of A2029.  We choose tCDM as the background cosmology.
A critical universe has larger accretion rates at $z=0$, and then
smaller shock radius, with respect to the $\Lambda$CDM case (in a
$\Lambda$CDM universe, a shock radius $\simeq 20 $ \% larger than the
virial one is expected, see TN, and Figure \ref{fig4}).  The resulting
outputs are spatially and spectrally analyzed with XSELECT and XSPEC,
assuming an absorbed Raymond-Smith spectrum.  Background counts are
subtracted for all spatial data bins using a simulated observation of
blank sky, to account for vignetting effects. The spectral fits are
performed in concentric annuli, centred on the cluster core.  The
neutral hydrogen column density is fixed ($N_H = 3\times 10^{20}$
cm$^{-2}$ to match the value at A2029) and the redshift is fixed,
while the temperature, normalization and metallicity are allowed to
vary.  The metallicity is always poorly constrained in these
simulations.

\begin{figure}
\centerline{\psfig{figure=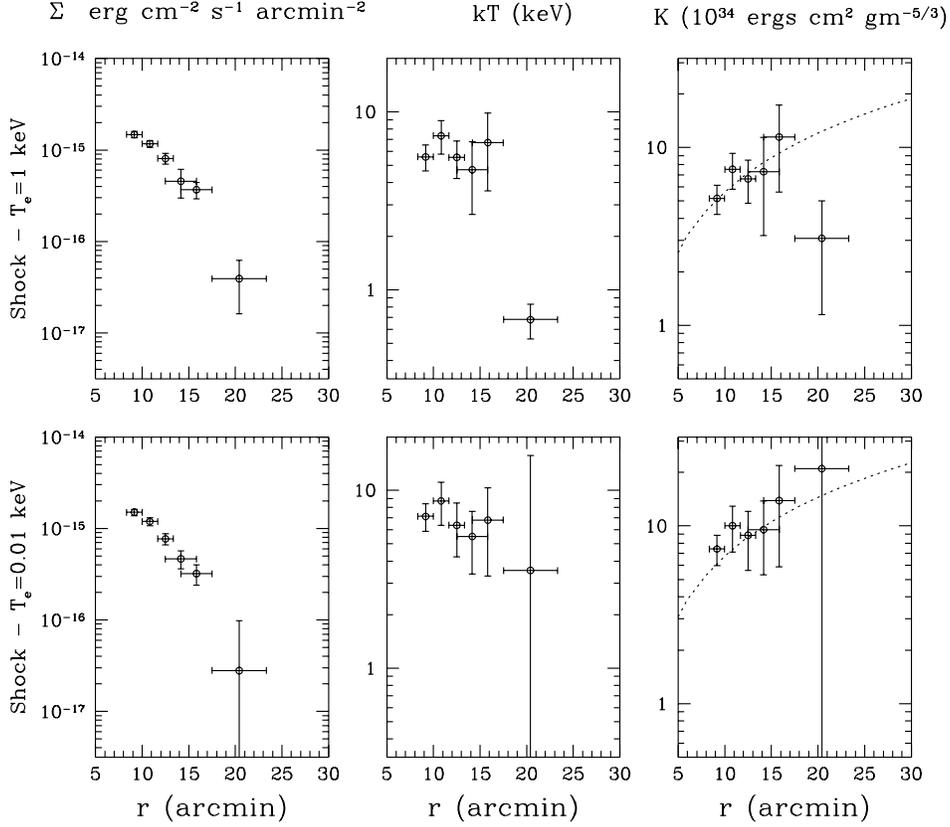,height=5.in}}
\caption{Simulated observations based on Abell 2029 for the two cases
discussed in the text.  The three columns show the observed surface
brightness, temperature and entropy profiles, with relative error bars
($90$ \% confidence level).  A beta-model with $\beta \simeq 0.7$,
normalized to the results of Sarazin et al. (1998), has been used.
First row: strong shock with an external warm gas ($kT_{e}=1$); second
row: strong shock with an external cold gas ($kT_e=0.01$).  The dotted
line is the power law
$r^{1.1}$ predicted for the shocked gas interior to the shock radius.
\label{fig5}}
\end{figure}

\begin{figure}
\centerline{\psfig{figure=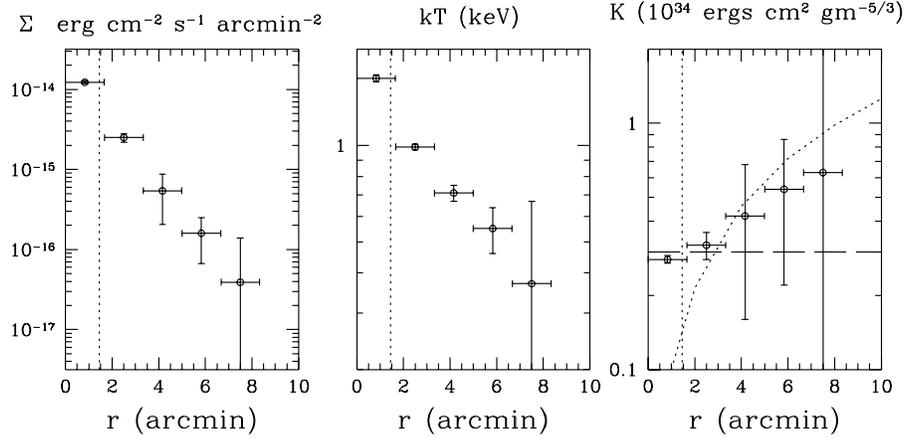,height=5.in}}
\caption{Simulated observation of a low mass group at the redshift of
A2029, exhibiting an adiabatic profile with no shocks. Note change in
x-axis scale from Figure \ref{fig5}. Vertical lines indicate the $100$
$h^{-1}$ kpc scale of current group luminosity measurements. In panel
3 a horizontal line is drawn at the constant (external) entropy level
of $K_{34}=0.3$. Although the entropy is poorly constrained at large
radii an inner entropy core is clearly observed ($r<3$ arcmin) and the
outer entropy is consistent with these values.
\label{fig6}}
\end{figure}

The simulated cluster observations are shown in Figure \ref{fig5}, where
only
the data of the PN detector have been used.  In a more realistic
observation, the use of the two MOS detectors significantly aids obtaining
a
stronger signal, or can decrease the required exposure time (see
Figure \ref{fig6}).  The errors in the figure correspond to 1-sigma.
In the first case, the external gas is detected and its temperature
measured with about $20$\% uncertainty (1 sigma).  The reason for the
small error in the external temperature, despite the low emissivity of
the external gas, is due to its value $kT_e \approx 0.7$; for this
value the exponential cutoff, from which the temperature is measured,
falls in the spectral region of maximum sensitivity for XMM.  In the
third panel the resulting entropy profile is shown.  Despite the large
uncertainty in the external value, the discontinuity in the entropy is
visible.  We point out that we assumed an external density profile
$\rho_{ext} \propto r^{-2}$.  The presence of substructures, such as small
clumps being accreted by the cluster, can make the
gas much more
visible in emission, due to the enhanced density.  Moreover, the
entropy of the gas is not changed by the presence of substructure,
which contributes only with adiabatic compression.  Therefore we
believe that the case shown in Figure \ref{fig5} is to be considered
realistic, if not pessimistic.

The second cluster case (Figure \ref{fig5}) envisages a strong shock front
with a cold,
low-emission external plasma ($kT_e<0.1$ keV), corresponding to the
case with $K_{34}\leq 0.3$ constant (here in a
$\Lambda$CDM; c.f. 2nd row in Figure \ref{fig2}).  In this
case only the gas internal to the shock radius is detectable, due to
the low entropy of the external gas.  The non-detection of an external
gas halo would not provide any direct constraint on the external
entropy level.  Note that the region of the cluster being observed
here is at a radius approximately twice that of the last significant
point of Sarazin et al. (1998): the surface brightness profile in
clusters has never been tested to such large radii.

Finally, the third case (Figure \ref{fig6}) is expected to represent
lower mass groups: pure adiabatically compressed ICM with a flat
entropy profile, and a corresponding steeper temperature gradient
following $T\propto \rho^{2/3}$.  The emission from the halo smoothly
fades into the external IGM, without any discontinuity.  In this case,
with $K_{34}=0.3$ (in $\Lambda$CDM), the surface brightness is
characterized by a relatively large core.  A very interesting aspect
of this observation is the detection of emission from small groups at
an unprecedented distance from the center.  At the same redshift of
A2029, 6 arcmins correspond to 0.4 $h^{-1} $ Mpc.  At such a radius,
the surface brightness and the temperature gradient are clearly
detected.  The entropy profile is flatter than the shocked power law
in the center, but the errors are too big at $0.4 h^{-1}$ Mpc to
discriminate between a shocked and and adiabatic profile in the very
external regions.

As already mentioned, such low surface brightness emission is now
emerging from ROSAT data (Helsdon \& Ponman 2000) and predicted by TN.
We recall that for practical reasons the current luminosities of loose
groups are
estimated only within a fixed radius of $100 h^{-1} $ kpc (see Ponman
et al. 1996), while the total luminosities can be higher by a factor
larger than 3 when including all the gas accreted from the halo.  Note
however, that here and in TN we define an {\sl accreted}  mass of
baryons $M_B = f_B M_{tot}$.  In the case with $K_*\geq 0.1$, the most
external part of this gas is compressed (but not shocked) by the
presence of the potential well and can be located at radii as large as
three times the virial radius, thus it can hardly be said ``to be
accreted''.  This reflects the difficulty in defining X--ray emission
in small mass halos, in contrast to large mass halos where the X--ray
emission is dominated by the central regions.  These considerations
add further interest to tracing the X--ray emission of small mass
objects to the largest radii.

In Figure \ref{fig6} we show, as a function of the total X-ray
luminosity, the exposure times needed with XMM to detect the emission
interior and exterior to the accretion shock with a signal to noise of
$5$, {\em and} enough photons to derive the temperature to within 20\%
uncertainty.  Here we use the $2-10$ keV luminosity, and map to $T$
using the 2-10 keV EXOSAT $L$-$T$ relation (David et al. 1993).  We
have assumed the same redshift as Abell 2029 ($z= 0.0767$),
$\beta\simeq 0.7$, and an external temperature of about $1$ keV.  As
we already discussed, such an external temperature is expected in the
case with $K_{34} = 3 \, (1+z)^{-2}$ , while a constant $K_{34}\simeq
0.2-0.4$ would give an almost undetectable external gas.  Thus we use
Figure \ref{fig6} as a guide to the needed exposure time in the
cases when the shock is detectable (i.e., $kT > 0.2$ keV); we recall
that temperatures lower than 1 keV (but still larger than 0.2 keV) are
more easily constrained due to their stronger exponential cutoff.  The
limits
in Figure \ref{fig6} are derived using the signal in the PN + 2MOS
detectors, for different choices of the ratio $R_S/R_V$ of the shock
to the virial radius (as long as $\rho_i/\rho_e > 1$).  We have always
assumed the shock fronts to be at the center of the XMM field of view.
The small circle represents our simulated observation of Abell
2029. The constraints on the observation times are dominated by the
requirement to have $4000$ and $2000$ photons respectively inside and
outside the shock (continuous lines) while the requirement of the
5-sigma emission detection becomes dominant at lower luminosities
(dashed lines).  It is clear that, with sufficent exposure at lower
luminosities, the shock/adiabatic transition can be mapped to a
considerable extent, allowing a direct test of the general picture
summarized in \S 2.

Looking further ahead, two possible missions would make the cluster
accretion shock a routine observation in studies of clusters.  ESA's
X-ray evolving Universe Spectroscopy (XEUS) mission, has design goals
for a $3\times 10^5$ cm$^{2}$ effective area ($\sim 70$ times larger
than XMM) and sub 2-arcsec imaging, with high spectral
resolution. NASA's CONSTELLATION-X with a factor 20-100 times larger
area than current missions, plus the ability to perform
high-resolution spectroscopy of extended objects, could potentially
see line emission from pre-shock gas superimposed on the continuum
emission of the shocked gas.

\section{Discussion}

From empirical evidence it seems clear that non--gravitational heating
plays a
key role in the thermodynamics of the ICM, but the physical mechanism
responsible for this heating is not known.  A debate exists over 
whether the sources of heating are mostly stars or AGN (see Wu \&
Fabian 1999; Valageas \& Silk 1999; Menci \& Cavaliere 2000;
Loewenstein 2000).  Numerical simulations that try to include stellar
feedback and cooling still have difficulties in linking the small
scale physics to the large scale dynamics of the ICM.  Our approach
 starts simply from the analysis of the thermodynamics of the ICM
as it is seen in local, and possibly distant clusters, and attempts to
answer the most direct questions.  We do not  build a
heating scenario a priori, rather we want to investigate possible
scenarios by directly studying the thermodynamic state of the ICM.  A key
question that we address here is what if the gas has been heated before or
after the collapse of the X--ray halo, or, in other terms, at low
(with a relatively small energy input) or at large densities (with a
larger energy budget)?  Here we briefly review the consequences for a
scenario where stars provide most of the heating energy.

As described in \S 1, the entropy level measured in high--$z$
Ly$\alpha$ clouds is low with respect to the level observed in
clusters of galaxies.  From Figure 10b of Ricotti et al. 2000, we can
interpolate $K_{Ly_\alpha}\sim 1.6 \, 10^{-2} (1+z)^{-1}\times
10^{34}$ erg g$^{-5/3}$ cm$^2$; higher values of this entropy would
make the IGM invisible in absorption.  Thus, we estimate that the
ratio of the entropy $K_{cl}$ observed in the clusters to that
observed in Ly$\alpha$ is $K_{cl}/K_{Ly_\alpha}\simeq 10 (1+z)$.  This
difference is even larger for the higher values of $K_*$ that enable a
good fit to the local $L$--$T$ relation and are still allowed by
current data.

If we assume that the gas seen in the Ly$\alpha$ clouds is
representative of the majority of the IGM, we are witnessing a clear
evolution in the equation of state of the diffuse baryons going from
the low densities of the Ly$\alpha$ clouds to the higher densities of
the X--ray halos.  In the framework of the entropy model, this
indicates that the IGM undergoes substantial heating just before or
after being accreted into the potential wells of groups and clusters.
Additionally, the chemical properties of the
IGM seen in the Ly$\alpha$ forest are different from those of the
ICM in clusters, indicating that the ICM is affected by star
formation processes and chemical enrichment, with a commensurate
amount of entropy production.  The subsequent questions are: when does
this
heating occur?  Can the star formation processes do the job?  To
determine if star formation processes can be solely responsible for
the excess entropy or just provide a minor contribution, a clear
correlation between the epoch of star formation and that of the excess
entropy production must be established.

\begin{figure}
\centerline{\psfig{figure=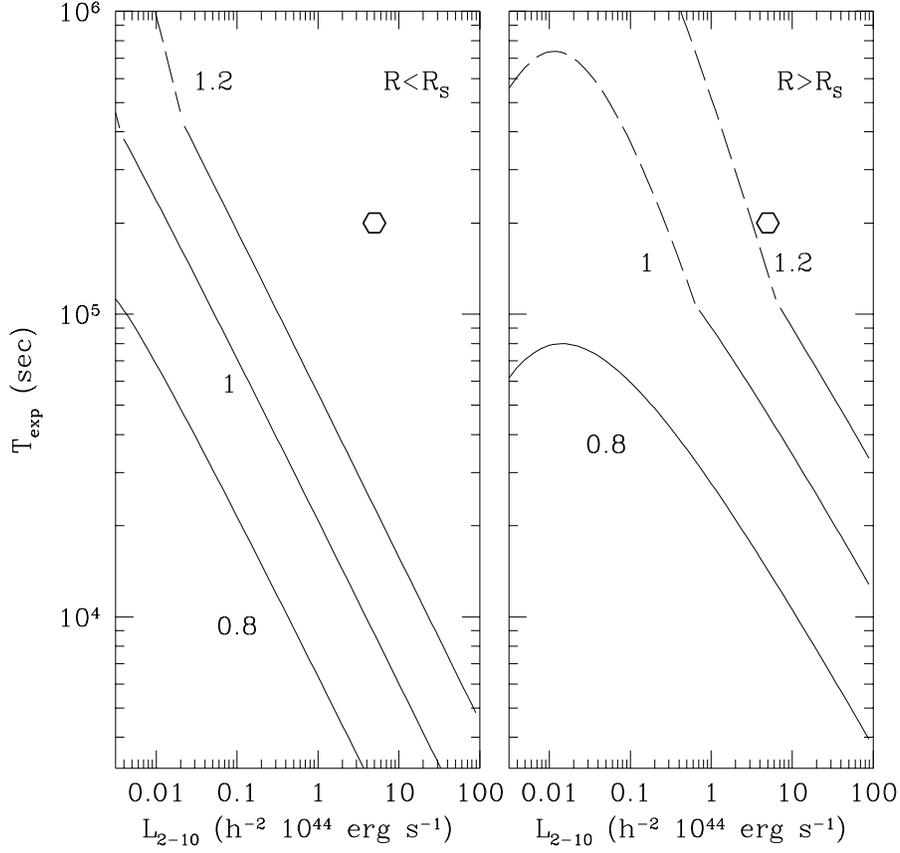,height=5.in}}
\caption{ The exposure times needed to detect the emission internal
(left panel) and external (right panel) to the shock, with a S/N$= 5$,
as a function of the total $2-10$ keV luminosity for objects at the
same redshift of Abell 2029 ($z= 0.0767$).  Here $\beta =2/3$ and
$kT_e=1$ keV.  The limits are derived using the signal in the PN+2MOS
detectors (thin filter).  Different curves refer to different value of
the ratio $R_S/R_V$.  The circle refers to the simulated observations
for Abell 2029. 
\label{fig7}}
\end{figure}

From the entropy profiles of local clusters we can extract  
temporal information. We predict a major feature of these profiles to be
the transition between a shock induced power law and a central
(adiabatic) entropy
core.  When the entropy plateau is eroded by cooling, especially in
larger halos, the central entropy level can no longer be directly related
to
the initial $K_*$ value.  However, the transition between the shock
and the adiabatic regime is still a relevant and robust feature,
marked by a change of slope in the entropy profile.  Meaningful
quantities are this transition radius $r_{t}$, and the baryonic mass
enclosed
within the radius itself, $M_{ad}$.  These two quantities, in the
external heating scenario, are almost constant between clusters and
groups,
with a clear dependence on the parameter $K_*$ (see Figure
\ref{fig7}). The baryonic fraction, of course, refers only to the
diffuse, hot gas, and not to  gas that may have cooled and sunk to the
center.  So, as a first approximation, the measure of $M_{ad}$ or
$r_t$ at several mass scales will provide a test of the simplest
external scenario with a single value for $K_*$, and at the same time
an indication on the level of $K_*$ itself.  Moreover, in larger
halos, the internal adiabatic core has been accreted at higher
redshift (third panel of Figure \ref{fig7}).  Detecting the presence
of the adiabatic transition in large halos, will put a lower limit to
the redshift when the entropy $K_*$ must be already in place.

In the internal case, a transition radius is not defined.  Rather, the
non--gravitational entropy contribution is simply superimposed on the
shocked profile.  We do not consider a physical
model for the internal heating, and thus we do not make specific
prediction for the corresponding entropy profile.  Nevertheless, the
detection of a break in the entropy profile, together with a constant
baryonic mass enclosed within $r_t$, would favour the external
scenario.  In this way the measure of $M_{ad}$ and $r_t$ can
significantly constrain this model and probe if the stellar populations 
provide
the bulk of the excess entropy.  These pieces of information can be
combined with other measures from different wavelenghts to better
evaluate the contribution of star formation processes to the global
heating.

The final part of this discussion is devoted to the effect of
substructures in the infalling medium.  Some of the baryons
can be shocked in small sub--halos before being accreted by the main
progenitor.  Such a gravitationally--produced entropy would raise
the average entropy level around large clusters of galaxies, with
respect to the average value $K_*$ in the non--shocked gas.  However,
the entropy which is generated in these gravitational processes does not
break the self similarity, since it always scales with
the mass of the accreting halo.  In other words, shock heating
processes such as these are not able to generate an entropy plateau in the
center
of X--ray halos. Moreover, in the presence of a
minimum entropy $K_*$, such external gravitational contribution
rapidly vanishes at small scales, since the satellites of smaller
halos are correspondingly smaller and unable to shock the baryons.
Thus, the net effect of a moderate amount of substructure around halos is
to enhance the detectability of the gas without changing its entropy.

In the present treatment we are not including the {\sl bow}
shocks developed through the merging of cluster subunits of comparable
mass (indicated by {\it ASCA} and {\it ROSAT} observations, cf.
Henriksen \& Markevitch 1996, Donnelly et al. 1998). In this
case strong non-equilibrium features appear (hot spots) and the plasma
is vigorously stirred.  However, the occurrence of large, violent
mergers is expected to be relatively rare in (for example) Cold Dark
Matter dominated cosmologies within the framework of the
extended Press-Schechter theory.  Most of the mass growth of a typical
cluster occurs by accretion of small clumps and diffuse matter onto a
main progenitor, the relative amounts of which depend on the details
of the cosmology and mass power spectrum. In most CDM models,
dynamically {\sl quiet} clusters always constitute a significant
fraction of the total population, especially at $z=0$. It is these
systems that are most likely to exhibit well defined accretion shocks.
The picture is different at redshift $z\simeq 1$, where the accretion
rate and then the rate of massive merger are about an order of
magnitude larger with respect to $z=0$.  We estimate (from the
extended PS theory) that the average number of major mergers (i.e.,
with a mass ratio larger than $0.3$) occurring within $1$ Gyr, is $<0.1$
at $z=0$ and $0.3-0.6$ at $z=1$ in a $\Lambda$CDM
cosmology.  Thus the
fraction of X--ray halos possibly affected by massive merger increases
dramatically at high redshifts.  Such a population of halos would ideally 
be modelled with hydrodynamical simulations, which can
capture the full three dimensional complexity of the processes.

\section{Conclusions}

In this work we have described how to investigate the thermodynamics
of the intra-cluster medium by resolving the entropy distribution
within X--ray halos.  The ability to spatially and spectrally resolve
nearby groups and clusters ($z\leq 0.1$) with current X--ray
satellites, can provide many crucial observations: the measure of the
entropy level of the non--shocked gas at $z\simeq 0$ around clusters
of galaxies; the detection of the extended, low surface brightness
emission at large radii in groups, and the transition from the adiabatic
to the shock regime imprinted in the inner entropy profile of X--ray
halos (expected at about $r_t\simeq 0.2-0.4 h^{-1} $ Mpc in the
external scenario).  Such observations will help in probing the two
basic scenarios adopted here to describe the non--gravitational
heating of the ICM:  {\sl external} and {\sl internal}.

In the simpler case of the external scenario, the only free parameter
is the entropy excess $K_*$ initially present in the diffuse IGM.  In
this case,  entropy profiles will statistically constrain the value
of $K_*$ and can be used to put a lower limit on the redshift of the
heating.  This will allow an estimation of the epoch and energetics of
the heating process itself, and will help to answer  the question of
whether or not the star formation processes are responsible for the bulk
of the
entropy excess.

One of the most exciting possibilities is the direct measurement of the
external entropy from the emission of the accreting IGM just outside the
shock radius of very massive clusters.  

In detail, the detection of the external entropy of the
pre--shocked gas requires a measurement of both the surface brightness
and temperature of cluster gas around the shock radius.  With such
data, the entropy profile across the shock can be derived, and hence
the thermodynamic state of both the ICM {\em and} the IGM. The
detection of accretion shock signatures in rich clusters, together
with the observation of constant entropy profiles in groups, would be
consistent with the hypothesis of an excess entropy in the external
IGM, accreted by dark matter halos. In particular, if the measured
entropy level in the gas around clusters and in groups is similar, the
simple scenario of homogeneous entropy production in the IGM at high
redshifts will be strongly supported.  This would simultaneously help
constrain physical models for the generation of the entropy.

We describe simulated observations of clusters and groups
with XMM to assess feasibility.  For two representative values
assumed in the external entropy, we show how to detect the
low--surface brightness gas at large radii both in large and small
halos.  In particular, in large halos a discontinuity in the entropy
may be visible, corresponding to the shock radius, while in 
small halos, a continuous isentropic distribution is expected, possibly
extending to very large radii.

A failure to detect the excess entropy in outer, non--shocked gas in
massive clusters, would favour the internal scenario, in which the excess
entropy is produced within the X--ray halo after the accretion, thus when
the gas has already reached higher densities.  In this case the energy
budget required to attain the same entropy excess at $z=0$ is much higher
(more than $1$ keV, with respect to the $\simeq 0.1$ keV required in the
external scenario).  Moreover, the internal scenario may leave an imprint
in the {\sl internal} entropy profile of X--ray halos which is at variance
with the profiles $K\propto r^{1.1}$ expected in the external scenario.  
Indeed, the capabilities of current X-ray satellites may be sufficient to
 image the structure of enhanced
internal entropy production.

Such observations would therefore provide crucial information at the
confluence of many different physical processes involving both baryons
and dark matter, that put in a common perspective an enormous amount
of data, both in the optical and the X-ray band.  At present there are
no other viable observations which can connect the entropy of the IGM
detected, e.g., in the Ly$\alpha$ forest with the entropy level
required to explain X--ray constraints from galaxy clusters and
groups.  We show how an instrument such as XMM can relatively easily
perform the necessary measurements and hope this work encourages
future observations which will directly test the cluster physics
described here.

\acknowledgements We thank Megan Donahue and David Strickland for
their help in the use of XSELECT and XSPEC.  We acknowledge
interesting discussion with the participants in the Milano 1999
workshop on ``Evolution of Galaxies in Clusters'', especially
A. Babul, R. Bower, and N. Menci.  We thank also the anonymous referee
for stimulating comments.  This work has been supported by NASA grants
NAG 8-1133. CAS acknowledges the support of NASA grant NAG 5-3257.

\newpage

\begin{figure}
\centerline{\psfig{figure=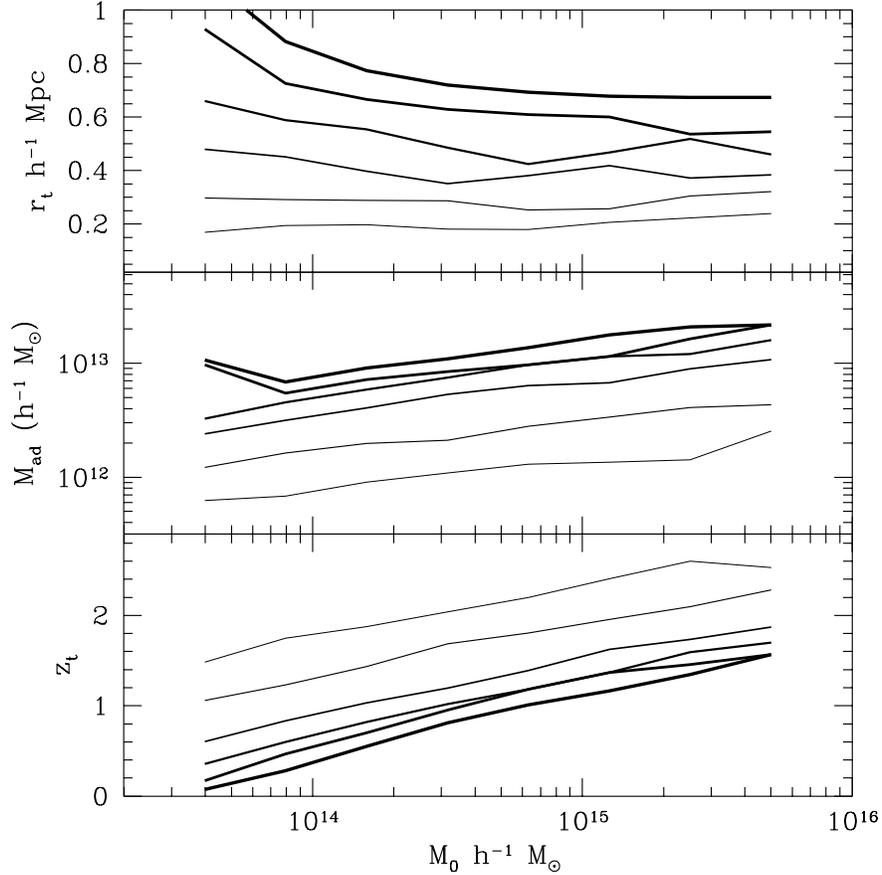,height=5.in}}
\caption{The transition radius, the mass included in the transition
radius, and the epoch of the adiabatic/shock transition as a function
of the total mass scale for the external scenario.  The curves
correspond to different values of $K_*$, from heaviest lines to
lightest: 0.5, 0.4, 0.3, 0.2, 0.1, and $0.05 \times 10^{34}$ erg
cm$^2$ g$^{-5/3}$.  A $\Lambda$CDM cosmology has been adopted.
\label{fig8}}
\end{figure}

\end{document}